\documentclass[%
 aip,
 app,
 amsmath,amssymb,
 reprint,
]{revtex4-2}

\newcommand{\laosto}{LaAlO$_3$/SrTiO$_3$ }

\newcommand{\sto}{SrTiO$_3$ }
\usepackage{graphicx}
\usepackage{xcolor}
\usepackage{dcolumn}
\usepackage{bm}
\usepackage[utf8]{inputenc}
\usepackage[T1]{fontenc}
\usepackage{mathptmx}
\usepackage{etoolbox}
\usepackage{amsmath}

\makeatletter
\def\@email#1#2{%
 \endgroup
 \patchcmd{\titleblock@produce}
  {\frontmatter@RRAPformat}
  {\frontmatter@RRAPformat{\produce@RRAP{*#1\href{mailto:#2}{#2}}}\frontmatter@RRAPformat}
  {}{}
}%
\makeatother

\begin{document}

\preprint{}

\title{Electric-Field-Induced Second Harmonic Generation at a
Reconfigurable \laosto Nanojunction}

\author{Pubudu Wijesinghe}
\affiliation{Department of Physics and Astronomy, University of Pittsburgh,
Pittsburgh, Pennsylvania 15260, USA}
\author{Melanie Dieterlen}
\affiliation{Department of Physics and Astronomy, University of Pittsburgh,
Pittsburgh, Pennsylvania 15260, USA}
\author{Kyoungjun Lee}
\affiliation{Department of Materials Science and Engineering,
University of Wisconsin-Madison, Madison, Wisconsin 53706, USA}
\author{Ahmed Omran}
\affiliation{Department of Physics and Astronomy, University of Pittsburgh,
Pittsburgh, Pennsylvania 15260, USA}
\author{Aswini Ramankutty}
\affiliation{Department of Physics and Astronomy, University of Pittsburgh,
Pittsburgh, Pennsylvania 15260, USA}
\author{Chang-Beom Eom}
\affiliation{Department of Materials Science and Engineering,
University of Wisconsin-Madison, Madison, Wisconsin 53706, USA}
\author{Patrick Irvin}
\author{Jeremy Levy}
\email{jlevy@pitt.edu}
\affiliation{Department of Physics and Astronomy, University of Pittsburgh,
Pittsburgh, Pennsylvania 15260, USA}

\date{\today}

\begin{abstract}
Electrically tunable nonlinear optical responses at the nanoscale remain challenging to achieve because conventional nonlinear materials lack the combination of large susceptibility, nanoscale confinement, and \textit{in situ} reconfigurability. Here we report electric-field-induced second harmonic (EFISH) generation from a nanoscale tunnel junction defined by conductive atomic force microscope lithography at the \laosto interface. A conducting channel written at the interface is interrupted by a nanoscale insulating gap, across which applied DC bias produces local electric fields exceeding $10^7$~V/m. The SHG signal is spatially localized at the junction, exhibits a quadratic bias dependence described by $I(2\omega) \propto |\chi^{(2)}_\mathrm{0} + \chi^{(3)} E_\mathrm{DC}|^2$ with no hysteresis, a modulation depth exceeding 380\% at $|V_\mathrm{DC}| = 1$~V, and shows a two-lobed input-polarization pattern aligned with the junction axis, consistent with EFISH from a centrosymmetric host. Calibration against a BBO reference crystal gives $|\chi^{(3)}| \approx 1\times10^{-19}$~m$^2$/V$^2$ at 6~K. These results establish cAFM-written oxide nanojunctions as a reconfigurable platform for nanoscale nonlinear optics in which the junction geometry sets the symmetry of the response and the large field-induced $\chi^{(2)}$ of \sto\ provides the optical nonlinearity. Because the nonlinearity is both generated and read out within the same nanoscale gap, the junction operates simultaneously as a subwavelength source and a near-field detector of optical nonlinearity.
\end{abstract}

\maketitle

The ability to control nonlinear optical interactions at the nanoscale is central to integrated photonics, ultrafast sensing, and cryogenic quantum electro-optics, where electro-optic modulation and microwave-to-optical transduction call for large nonlinearities at low temperature~\cite{Boyd2003NonlinearOptics,Wang2022-bt}. Electric-field-induced second harmonic (EFISH) generation is particularly attractive because it enables tunable, field-induced second-order responses in centrosymmetric materials whose bulk symmetry would otherwise forbid such processes~\cite{Terhune1962-cd,Corn1994-bf}: an applied DC field breaks inversion symmetry locally, inducing a second-order polarization proportional to $\chi^{(3)} E_\mathrm{DC}$ that scales with both the medium's third-order susceptibility and the applied field~\cite{Bloembergen1968-je}. Large EFISH signals therefore require materials with exceptionally large $\chi^{(3)}$, high local fields, and well-defined symmetry-breaking geometry.

Oxide heterostructures are emerging as a compelling platform for interfacial nonlinear optics. Recent work demonstrated EFISH from polar vortex structures in PbTiO$_3$/SrTiO$_3$ superlattices, where electric-field-driven manipulation of polar skyrmions modulated the SHG intensity by hundreds of percent~\cite{Wang2024-sr}. Electric-field-controlled SHG has also been reported in two-dimensional materials~\cite{Shi2019-ed}, silicon photonic devices~\cite{Timurdogan2017-bp}, and ferroelectric thin films~\cite{Fiebig2005-fn}, but these platforms typically lack nanoscale spatial control and \textit{in situ} reconfigurability of the nonlinear source geometry.

The LaAlO$_3$/SrTiO$_3$ (LAO/STO) interface hosts a two-dimensional electron gas (2DEG) that arises above a critical LAO thickness of four unit cells~\cite{Ohtomo2004-ed,Thiel2006-ds}, and can be locally switched between conducting and insulating states by voltages applied through a conductive atomic force microscope (cAFM) tip~\cite{Cen2008-rg,Cen2009-ja}. This method enables the creation and erasure of conducting nanowires, tunnel junctions, and single-electron transistors with nanometer-scale precision~\cite{Bi2010-by,Irvin2010Rewritable,Cheng2011-ea}; the same region can be written, erased, and rewritten many times~\cite{Cen2009-ja}. No other complex oxide platform offers this combination of nanoscale spatial control and reversible reconfigurability.

SrTiO$_3$ is a quantum paraelectric: its dielectric constant rises dramatically upon cooling, following a Curie-Weiss-like divergence that saturates near 35–40 K instead of diverging, due to quantum fluctuations, leaving STO in a paraelectric state at all temperatures~\cite{Muller1979-wa}, with $\varepsilon_r$ reaching values of $10^3$--$10^4$ below 100~K~\cite{Hemberger1995-el}. This large static susceptibility, the response of the soft polar transverse-optical (TO) mode, is what makes STO an efficient EFISH medium (mechanism detailed below); the relevant $\chi^{(3)}$ enters through this quasi-static soft-mode channel and so scales with the static ($\omega \rightarrow 0 $) susceptibility $\varepsilon_s (T,E)$ rather than with the modest electronic permittivity $\varepsilon_\infty \approx 5$--$6$ sampled by the optical fields themselves~\cite{Hemberger1995-el,Hemberger1996-iw}. Because $\varepsilon_s(T,E)$ saturates under a large applied field, the nonlinearity is large but its low-temperature enhancement is bounded rather than divergent, as we establish below.

A central motivation for this work is to establish the optical character of the nonlinear response of cAFM-defined LAO/STO junctions, which for more than a decade have served as nanoscale nonlinear-optical elements that both generate and detect light: as broadband terahertz emitters and detectors~\cite{Ma2013-hv,Chen2019-hz}, as rewritable nanoscale photodetectors~\cite{Irvin2010Rewritable}, and as near-field probes of co-located nano-objects~\cite{Jnawali2015-un,Sheridan2021-jg}. Because optical ($\chi^{(3)}$) and electrical nonlinearities can produce a similar dependence on bias, the optical origin of this response has been hard to establish from transport alone. Detecting the sum frequency resolves this directly: the second-harmonic signal reported here is the sum-frequency counterpart of the bias-mediated difference-frequency mixing through which the same $\chi^{(3)}$ generates terahertz radiation at these junctions~\cite{Chen2019-hz}, confirming a genuinely optical, field-tunable element.

Here we combine these two features, cAFM-defined nanojunctions and the large field-induced $\chi^{(3)}$ of STO, to demonstrate EFISH from a reconfigurable nanoscale source. We write a conducting channel at the LAO/STO interface, interrupt it with a nanoscale insulating gap to form a tunnel junction, and apply a DC bias across the gap. The resulting SHG signal is localized to the junction and tunable in amplitude and sign by the applied bias. Because generation and detection are reciprocal, the same $\chi^{(3)}$ element that emits the second harmonic is also sensitive to its immediate surroundings, so source and detector are co-located within a single, deeply subwavelength gap.

\begin{figure*}[t]
    \centering
    \includegraphics[width=\linewidth]{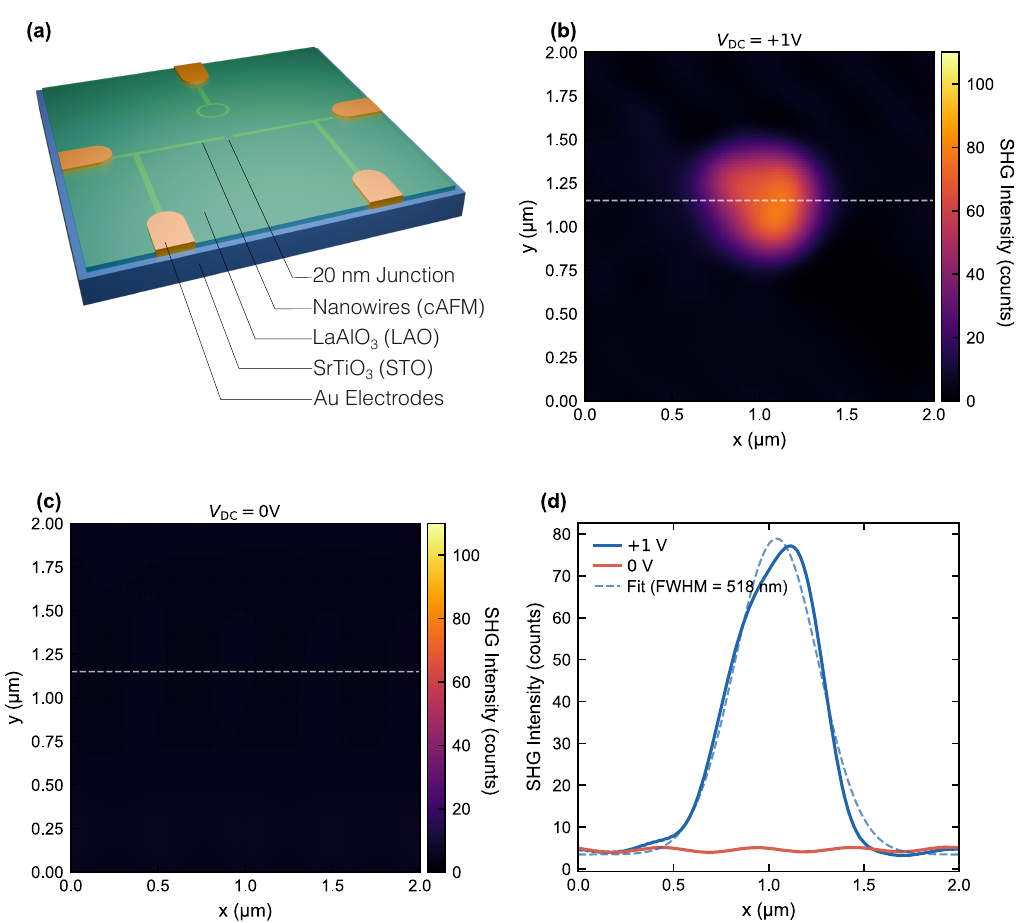}
\caption{Device geometry and spatial imaging of the EFISH response. (a)~Schematic of the cAFM-written LAO/STO nanojunction with gold contact pads; a conducting nanowire along $[100]$ is interrupted by a $\sim$20~nm insulating gap, with DC bias $V_\mathrm{DC}$ applied across the junction. (b) SHG map at $V_\mathrm{DC} = +1$~V, showing a localized enhancement at the nanogap ($\sim$80 counts vs.\ $\sim$5 counts background). (c) SHG map at zero bias, showing no localized signal above background. (d)~Line profiles for both bias conditions; the $+1$~V profile is fitted by a Gaussian (dashed) with FWHM~$\approx 500$~nm, near the diffraction limit for 400~nm light at NA~$= 0.5$.}
    \label{fig:device}
\end{figure*}

LAO/STO heterostructures (3.4~unit cells LAO) were grown by pulsed laser deposition on TiO$_2$-terminated SrTiO$_3$(001) substrates (see supplementary material). A conducting nanowire written by cAFM was interrupted by a nanoscale insulating gap ($\sim$20~nm) to form the junction; gold contact pads provide electrical access for applying $V_\mathrm{DC}$ across the gap (Fig.~\ref{fig:device}a). With the full voltage dropping across a $\sim$20~nm gap, the local electric field at $V_\mathrm{DC} = 1$~V reaches $\sim$$5 \times 10^7$~V/m, well into the regime where this response from the quantum paraelectric STO is expected to be significant~\cite{Hemberger1995-el,Anderson2025-ca}.
 
Upon application of $V_{\text{DC}} = +1\text{ V}$ (Fig. 1b), a pronounced localized enhancement appears at the nanogap. In contrast, wide-field SHG imaging at zero bias (Fig. 1c) reveals no localized feature at the junction location, indicating that the bias-independent background $I_0 = \vert{}\chi_{\text{0}}^{(2)}\vert{}^2$ lies below the detection limit in the absence of an applied field. The signal remains negligible along the conducting wire arms and the unwritten $\text{LaAlO}_3/\text{SrTiO}_3$ surface, confirming that the enhancement is strictly confined to the high-field junction region. 

The line profile through the junction (Fig.~\ref{fig:device}d) confines the SHG enhancement to a Gaussian spot with FWHM~$\approx 500$~nm, close to the diffraction limit, confirming that the nonlinear source is localized within the $\sim$20~nm junction gap, well below the optical resolution. This confinement follows directly from the junction geometry: the applied voltage drops entirely across the nanogap, concentrating the symmetry-breaking field at the point of interruption rather than along the extended wire. The same confinement makes the source sensitive to its immediate surroundings, so the junction acts as a near-field transducer of the local field and dielectric environment rather than merely a localized emitter.

SHG images were background subtracted and Fourier-filtered to remove periodic instrumental artifacts from scanning; the filter was applied identically to all images and does not affect spatial frequencies associated with the junction signal.

\begin{figure*}[t]
    \centering
    \includegraphics[width=\linewidth]{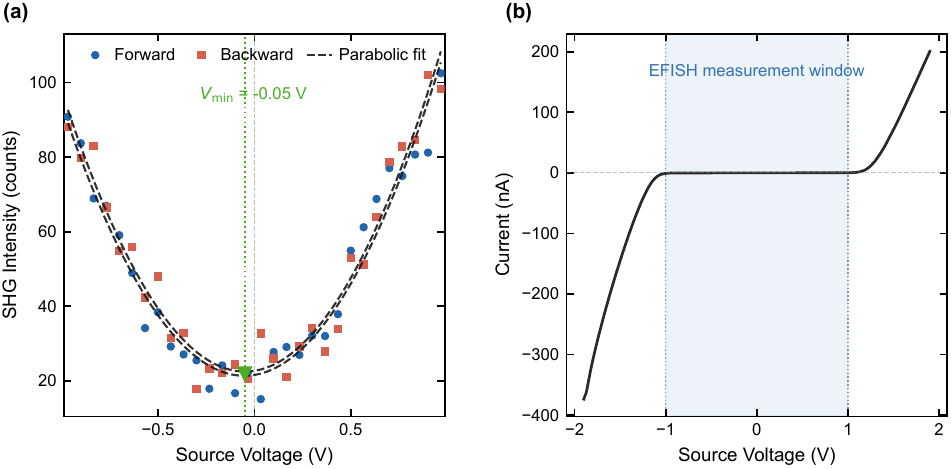}
\caption{Bias dependence of the EFISH signal at the LAO/STO nanojunction. (a)~SHG intensity vs.\ applied DC bias $V_\mathrm{DC}$, recorded in forward (blue) and reverse (red) sweeps. Solid curves are fits to Eq.~(\ref{eq:model_expanded}), yielding a minimum at $V_\mathrm{min} = -50 \pm 3$~mV consistent between sweeps; the finite signal at $V_\mathrm{DC}=0$ reflects the bias-independent $\chi^{(2)}_\mathrm{0}$ contribution, while the dominant quadratic term confirms the EFISH character of the response. (b)~Current--voltage characteristic over a wider range, confirming resistive behavior within the $\pm1$~V measurement window (shaded), consistent with negligible carrier transport during the optical measurement.}
    \label{fig:bias}
\end{figure*}

To characterize the nonlinear response, we measured the SHG intensity at the junction as a function of DC bias $V_\mathrm{DC}$ (Fig.~\ref{fig:bias}a). The signal rises from a finite background at $V_\mathrm{DC} = 0$ and increases with applied field in both polarities, with forward and reverse sweeps showing no hysteresis, confirming that no ferroelectric switching occurs and that the junction remains in the quantum paraelectric state throughout.

The general form of the SHG intensity at an interface with both a static symmetry-breaking contribution and a field-induced term is:
\begin{equation}
  I(2\omega) \propto \left| \chi^{(2)}_\mathrm{0}
  + \chi^{(3)} E_\mathrm{DC} \right|^2,
  \label{eq:model_full}
\end{equation}
where $\chi^{(2)}_\mathrm{0}$ arises from static inversion-symmetry breaking, whether structural and localized at the LAO/STO heterointerface~\cite{Rubano2023-ec} or intrinsic to the SrTiO$_3$ itself~\cite{Jang2010-so} and $\chi^{(3)}$ is the third-order susceptibility of STO; the $\chi^{(3)}$ relevant here is dominated by the soft TO mode through its quasi-static leg and so scales with the static dielectric susceptibility $\varepsilon(0,T,E)$ rather than with the electronic response, as discussed below~\cite{Fleury1968-mt, Hemberger1995-el}. Expanding Eq.~(\ref{eq:model_full}) gives:
\begin{equation}
  I(2\omega) = I_0
  + 2\chi^{(2)}_\mathrm{0}\chi^{(3)} E_\mathrm{DC}
  + |\chi^{(3)}|^2 E_\mathrm{DC}^2,
  \label{eq:model_expanded}
\end{equation}
where $I_0 = |\chi^{(2)}_\mathrm{0}|^2$ is the bias-independent SHG background from the interface, the linear term reflects interference between the two contributions, and the quadratic term is the pure EFISH response. Fitting Eq.~(\ref{eq:model_expanded}) as $I = a + bV_\mathrm{DC} + cV_\mathrm{DC}^2$ to the forward and reverse sweeps yields consistent parameters across both directions, with the signal minimum occurring at:
\begin{equation}
    V_\mathrm{min} = -\frac{b}{2c} = -50 \pm 3~\mathrm{mV}.
    \label{eq:vmin}
\end{equation}
Strictly, $V_\mathrm{min}$ measures $\mathrm{Re}(\chi^{(2)}_\mathrm{0}\chi^{(3)*})/|\chi^{(3)}|^2$; because $\chi^{(2)}_\mathrm{0}$ and $\chi^{(3)}$ need not be in phase (supplementary material, Sec.~S4.F), $|{-}b/2c|$ should be treated as a lower bound on $|\chi^{(2)}_\mathrm{0}/\chi^{(3)}|$ rather than an exact measurement. The dominance of the quadratic term ($c \gg b$) confirms that the response is overwhelmingly EFISH in character, while the nonzero $V_\mathrm{min}$ and finite $I_0$ are consistent with known structural symmetry breaking at the LAO/STO interface~\cite{Rubano2023-ec} and with intrinsic rectification from the cAFM cutting potential~\cite{Bogorin2010-tw}. At $|V_\mathrm{DC}| = 1$~V the bias-induced SHG intensity increases by $\sim$380\% relative to the zero-bias background ($\Delta I/I_0 = c/a$), demonstrating efficient field-induced tuning of the nonlinear response at the nanoscale.
The junction acts as a tunnel barrier within the measurement window, confirmed by the current--voltage characteristic (Fig.~\ref{fig:bias}b), ensuring the applied voltage drops across the junction without screening from carrier transport during the optical measurement. The SHG intensity scales with incident power as $I(2\omega) \propto P^{n}$ with $n = 1.89 \pm 0.08$ and $1.91 \pm 0.09$ at $V_\mathrm{DC} = +1$ and $-1$~V, consistent with a second-order process (see supplementary material).

To place this response on an absolute scale, we calibrated the quadratic (EFISH) term against a BBO reference crystal measured under matched excitation conditions (supplementary material, Sec.~S4). We estimate $|\chi^{(3)}| \approx 1\times10^{-19}$~m$^2$/V$^2$ at 6~K, roughly one to two orders of magnitude above typical room-temperature, non-resonant electronic (optical-Kerr) estimates for SrTiO$_3$ ($10^{-20}$--$10^{-21}$~m$^2$/V$^2$),~\cite{Fleury1968-mt} consistent with a modest soft-mode enhancement of the third-order response at low temperature and operating field (see supplementary material for the full calibration procedure, including a discussion of its systematic uncertainty).

\begin{figure*}[t]
    \centering
    \includegraphics[width=1\linewidth]{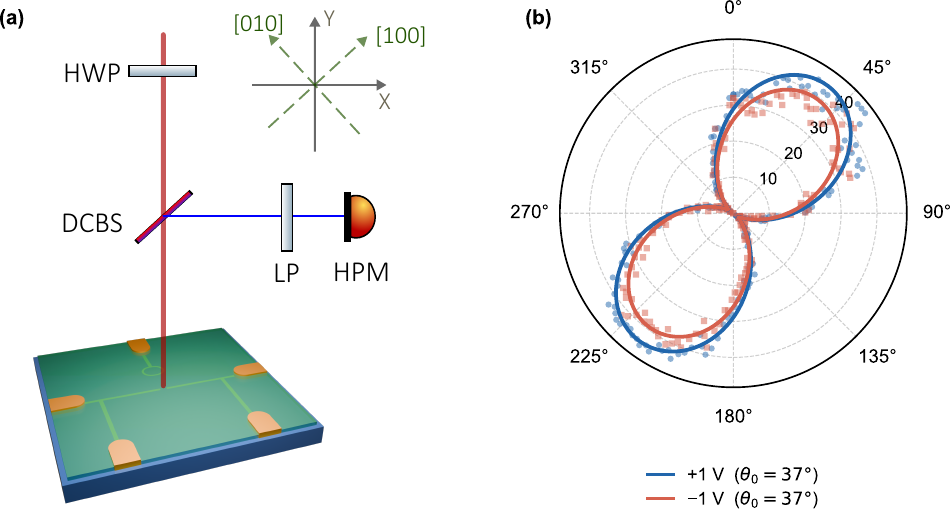}
\caption{\textbf{(a)} Schematic of the polarization-resolved EFISH measurement setup (see also supplementary material Fig. S1). The fundamental beam passes through a half-wave plate (HWP) to set the input polarization angle $\theta$, defined relative to the laboratory $Y$-axis, with the crystal $[100]$ and $[010]$ axes (dashed green) oriented at $\pm 45^\circ$. The signal is separated by a dichroic beam splitter (DCBS), filtered by a linear polarizer (LP), and measured with a hybrid photomultiplier (HPM). \textbf{(b)} Input polarization dependence of the EFISH signal from the LAO/STO nanojunction at $+1\text{ V}$ (blue) and $-1\text{ V}$ (red). Solid curves are $\cos^2$ fits yielding a lobe orientation $\theta_0 = 36.7^\circ$, consistent with the junction axis. The two-lobed pattern is characteristic of a field-induced response from a centrosymmetric medium; the amplitude difference between polarities ($A = 42.4$ vs. $37.6\text{ counts}$) reflects the nonzero $\chi_{\text{0}}^{(2)}$ at the interface.}
    \label{fig:polarization}
\end{figure*}

To determine the symmetry of the nonlinear source, we measured the SHG intensity as a function of the input polarization angle $\theta$ while keeping the analyzer fixed along the vertical direction. Figure~\ref{fig:polarization} shows the resulting polar plots at $V_\mathrm{DC} = +1$~V and $V_\mathrm{DC} = -1$~V.

Both sweeps exhibit a two-lobed pattern with lobes oriented along the junction axis ($\theta_0 = 36.7^\circ$), consistent with the cAFM writing direction along $[100]$ (at $45^\circ$ from the $Y$-axis in our laboratory frame); the $\sim$8$^\circ$ offset is attributable to mounting misalignment rather than any intrinsic rotation of the response. This is the expected signature for a uniaxial, field-induced source aligned with the applied field, with SHG maximized parallel to the junction axis and minimized perpendicular to it. Crucially, the lobe orientation tracks the junction geometry rather than the crystallographic axes of STO, which would instead produce a four-lobed pattern reflecting the $C_{4v}$ symmetry of the (001) surface~\cite{Zhao2013-ga} -- confirming that the symmetry breaking is set by the direction of $E_\mathrm{DC}$, not the underlying crystal structure.

The small but reproducible amplitude difference between $+1$~V and $-1$~V ($A = 42.4$ vs $37.6$~counts) at fixed lobe orientation is consistent with the interference term $2\chi^{(2)}_\mathrm{0}\chi^{(3)}E_\mathrm{DC}$ in Eq.~(\ref{eq:model_expanded}), which changes sign with bias polarity and modulates the total response asymmetrically.

\begin{figure}
    \centering
    \includegraphics[width=\linewidth]{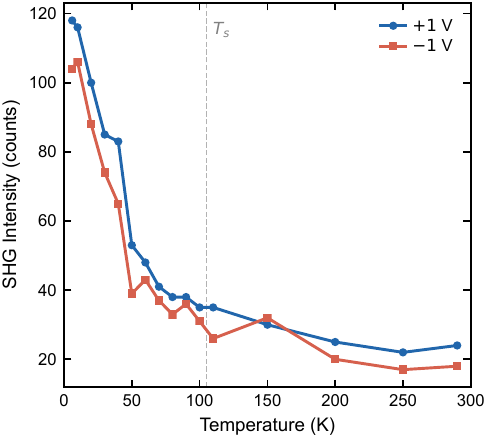}
\caption{Temperature dependence of the EFISH signal at the LAO/STO nanojunction at fixed $V_\mathrm{DC} = +1$~V (blue) and $-1$~V (red). The signal increases approximately fivefold upon cooling from 290~K to 6~K, a modest enhancement governed by the field-saturated quasi-static soft mode rather than the divergence of its small-signal permittivity (see text). The asymmetry between $\pm 1$~V persists across the full temperature range.}
    \label{fig:temp}
\end{figure}

The temperature dependence of this signal probes the soft-mode origin of the nonlinear response in STO. Figure~\ref{fig:temp} shows $I(2\omega)$ at fixed bias $V_\mathrm{DC} = \pm 1$~V as a function of temperature from 290 to 6~K. The signal is finite across the entire temperature range but rises sharply below $\sim$50~K, increasing approximately fivefold upon cooling to 6~K, and shows a tendency to plateau at the lowest temperatures.

In the idealized cubic paraelectric structure, every zone-center optic mode of STO has odd parity, so the second-order response is symmetry-forbidden; an applied DC field polarizes the soft TO mode, lowering the symmetry to $C_{4v}$ and switching on this response, the same mechanism established for electric-field-induced Raman scattering in STO and KTaO$_3$~\cite{Fleury1967-qf,Fleury1968-mt}. The DC field thus enters $\chi^{(3)}(2\omega;\omega,\omega,0)$ through a single quasi-static (soft-mode) leg, while the two optical legs at $\omega$ and $2\omega$ sample only the electronic permittivity $\varepsilon_\infty \approx 5$--$6$. Since the soft-mode response is the static susceptibility, with $\varepsilon(0) \propto \omega_\mathrm{TO}^{-2}$ by the Lyddane--Sachs--Teller relation, $\chi^{(3)} \propto \varepsilon(0,T,E)$ and $I(2\omega) \propto \varepsilon(0,T,E)^2\, E_\mathrm{DC}^2$~\cite{Fleury1967-qf,Hemberger1995-el}.

The small-signal $\varepsilon(0)$ of STO rises roughly a hundredfold between 290 and 6~K, which would naively imply a $\sim$$10^4$ enhancement of $I(2\omega)$ on cooling; this small-signal divergence, however, is not the relevant quantity at our operating field. At large DC bias the soft mode hardens and the static permittivity saturates as $\varepsilon(0,E) \propto E^{-2/3}$, with a quartic-anharmonic coefficient nearly temperature independent below $\sim$50~K~\cite{Worlock1967-bp,Akimov2000-pj}: at 8~K the soft-mode frequency already increases more than fourfold, and $\varepsilon(0,E)$ falls by an order of magnitude by only $\sim$1~MV/m~\cite{Worlock1967-bp}, so at $E_\mathrm{DC} \sim 5\times10^7$~V/m the mode is deep in saturation. The modest fivefold enhancement we observe is therefore the expected behavior of the saturated soft mode, not a failure of the underlying picture, and matches the same field-saturated quantum-paraelectric enhancement reported for STO as a cryogenic electro-optic material~\cite{Anderson2025-ca}.

No anomaly is observed at the structural transition temperature of STO ($T_s \approx 105$~K), where the cubic-to-tetragonal distortion breaks inversion symmetry and produces a distinct SHG feature in bulk STO~\cite{Rubano2023-ec}. The smooth, featureless evolution of our signal through 105~K confirms that the response is governed by the field-induced $\chi^{(3)}$ mechanism rather than the structural distortion of the substrate.

The bias-independent background $I_0$ need not arise solely from the
interfacial structural distortion discussed above. SrTiO$_3$ is an
incipient ferroelectric: its soft mode never condenses into long-range
polar order, but the material nonetheless hosts local, quenched polar
distortions, including oxygen-vacancy-related defect dipoles that have
been shown to produce a built-in polarization even in strain-free bulk
SrTiO$_3$~\cite{Jang2010-so}, and intrinsic polar fluctuations whose
correlation length grows on cooling as the soft mode approaches, but
never reaches, condensation. Either contribution would break inversion
symmetry locally and could add to $\chi^{(2)}_\mathrm{0}$ independent
of the LAO/STO interface. We do not attempt to separate these bulk and
interfacial contributions here; doing so would require, for example,
comparing junctions written on substrates with different defect
densities, or bare SrTiO$_3$ controls without the LAO overlayer.

The amplitude difference between $+1$~V and $-1$~V persists across the full temperature range, consistent with the interference term $2\chi^{(2)}_\mathrm{0}\chi^{(3)}E_\mathrm{DC}$ in Eq.~(\ref{eq:model_expanded}), which changes sign with bias polarity. Because this cross-term is small relative to the dominant quadratic response (supplementary material, Sec.~S4.F), our data do not resolve its temperature dependence. Since only the bias-induced $\chi^{(3)} E_\mathrm{DC}$ term carries the soft-mode (quasi-static) coupling, the bias and temperature dependences remain two views of the same underlying response, but disentangling the temperature dependence of $\chi^{(2)}_\mathrm{0}$ from that of $\chi^{(3)}$ will require additional measurements beyond the present data set.

Our results compare favorably with other reported EFISH platforms. Wang \textit{et al.}~\cite{Wang2024-sr} demonstrated a modulation of $\sim$664\% at an applied voltage of 7~V in PbTiO$_3$/SrTiO$_3$ polar skyrmion superlattices, operating at room temperature in a macroscopic device. Our junction reaches a comparable modulation depth, $\Delta I/I(0) \approx 380\%$, at $|V_\mathrm{DC}| = 1$~V, roughly seven times lower voltage, and in an active volume orders of magnitude smaller than the macroscopic superlattice. The nanoscale geometry also enables local electric fields exceeding $10^7$~V/m at modest applied voltages of a few volts, a regime inaccessible in bulk or thin-film geometries without dielectric breakdown. Furthermore, the full reconfigurability of the cAFM-written junction, the ability to erase and rewrite the nonlinear element at will, has no analog in any previously reported platform of this kind.

We have demonstrated EFISH from a reconfigurable nanoscale tunnel junction at the LAO/STO interface, establishing a new platform for nanoscale nonlinear optics in complex oxide heterostructures. The SHG signal is spatially localized at the cAFM-written junction, exhibits a parabolic bias dependence consistent with this mechanism in a centrosymmetric quantum paraelectric, and shows a polarization response that tracks the junction axis rather than the crystal symmetry, the hallmarks of a field-induced process. The temperature dependence is governed by the field-saturated quasi-static response of the soft mode rather than its divergent small-signal permittivity, with quantum paraelectricity enhancing the field-induced $\chi^{(3)}$ by $\sim 5\times$ without reaching the unconstrained bulk divergence.

These results open several directions. Because the junction is reconfigurable, its nonlinear source geometry can be set in situ via the cAFM writing direction, enabling orientation-selective polarimetry and, with on-chip waveguide integration, deeply subwavelength nonlinear elements for integrated photonics. The signal's sensitivity to the local field and dielectric environment also makes the device a local optical readout for cAFM-defined circuits and a probe for nanoscale electric-field sensing. The underlying soft-mode nonlinearity is the same resource sought for cryogenic electro-optic modulators and microwave-to-optical quantum transduction~\cite{Anderson2025-ca,Wang2022-bt}: the large field-tunable EFISH response implies a correspondingly large, voltage-tunable electro-optic response in a rewritable element, though soft-mode saturation bounds this below the giant small-signal values of STO near quantum criticality~\cite{Anderson2025-ca}. Arrays of such junctions, written, biased, and read both electrically and optically, would form a reconfigurable platform for nanoscale nonlinear optics, electro-optic modulation, and near-field sensing.

See supplementary material for further details on sample growth, cAFM device fabrication, the SHG measurement and calibration procedures, and the bias-application circuitry.

\begin{acknowledgments}
J.L. acknowledges support from ONR MURI N00014-21-1-2537, ONR N00014-20-1-2481, and the Department of Energy QIS program (DE-SC0022277). C.-B.E. acknowledges support from the Gordon and Betty Moore Foundation's EPiQS Initiative (Grant GBMF9065), a Vannevar Bush Faculty Fellowship (ONR N00014-20-1-2844) and the US National Science Foundation (NSF) through the Designing Materials to Revolutionize and Engineer our Future (DMREF) program (NSF award DMREF-2522669). Thin film synthesis at the University of Wisconsin-Madison was supported by the U.S. Department of Energy (DOE), Office of Science, Office of Basic Energy Sciences (BES), under award number DE-FG02-06ER46327.
\end{acknowledgments}

\section*{Data Availability Statement}
The data that support the findings of this study are available from the corresponding author upon reasonable request.

\clearpage
\onecolumngrid

\begin{center}
  \large \textbf{Supplementary Material for:}\\[0.4em]
  \Large \textbf{Electric-Field-Induced Second Harmonic Generation at a Reconfigurable \laosto Nanojunction}\\[1em]
\end{center}

\setcounter{section}{0}
\setcounter{figure}{0}
\setcounter{equation}{0}
\setcounter{table}{0}
\setcounter{page}{1}

\renewcommand{\thesection}{S\arabic{section}}
\renewcommand{\thefigure}{S\arabic{figure}}
\renewcommand{\theequation}{S\arabic{equation}}
\renewcommand{\thetable}{S\arabic{table}}
\renewcommand{\thepage}{S\arabic{page}}

\section{Bias Dependence}
During the DC bias sweeps, the applied voltage was ramped linearly in time from $-1$~V to $+1$~V over a total sweep duration of 2000~s. The SHG signal was recorded continuously via time-correlated single-photon counting (TCSPC) with a collection time of 2~s per point. Due to the discrete, low-photon-count nature of the signal, the raw time-series data is subject to Poisson noise.

To improve the signal-to-noise ratio and perform quantitative analysis, the photon counts were binned as a function of the applied DC bias voltage. The bias range was divided into 30 equally spaced intervals (bins). Within each bin, the mean photon count and the standard error of the mean (SEM), defined as $\text{SEM} = \sigma / \sqrt{N}$ (where $\sigma$ is the standard deviation of the counts and $N$ is the number of data points in the bin), were calculated. A weighted least-squares quadratic fit of the form $I(V_\mathrm{DC}) = a + b V_\mathrm{DC} + c V_\mathrm{DC}^2$ was then performed using the bin centers, with weights proportional to $1/\text{SEM}^2$.

For the symmetric sweep from $-1$~V to $+1$~V, this weighted fit yields parameters $a = 22$ counts (zero-bias background), $b = 8$ counts, and $c = 84$ counts.
The modulation depth, defined as the ratio of the quadratic coefficient to the zero-bias background, is $c/a = 3.81$, corresponding to a modulation depth of $\approx 380\%$ per $\text{V}$.

The parabolic fit also reveals a minimum at a nonzero voltage of $V_\mathrm{min} = -b/(2c) = -50 \pm 3$~mV. This built-in-field offset indicates the presence of a bias-independent, structural $\chi^{(2)}_\mathrm{0}$ contribution at the \laosto\ interface~\cite{Rubano2023-ec} and an asymmetry arising from the conductive atomic force microscope (cAFM) writing process, which acts as a built-in cutting-potential rectification at the written nanojunction~\cite{Bogorin2010-tw}.

For all bias sweeps reported here, the fundamental polarization was fixed parallel to the junction axis, i.e.\ along the crystallographic $[100]$ direction, coincident with the direction of $E_\mathrm{DC}$. This choice, made to maximize the collected SHG signal, has a direct consequence for the tensor analysis of Sec.~\ref{si:calibration}.

\section{Experimental Setup}
\begin{figure}
    \centering
    \includegraphics[width=0.5\linewidth]{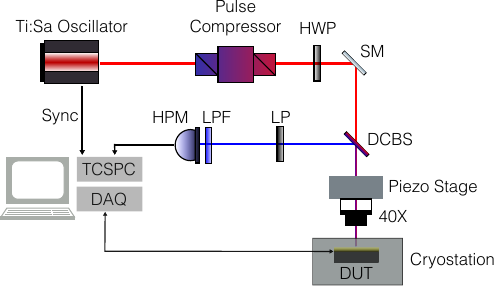}
    \caption{Schematic of the experimental setup for confocal electric-field-induced second harmonic (EFISH) microscopy at the reconfigurable \laosto\ nanojunction. A Ti:sapphire oscillator provides fundamental pulses at 800~nm, which are pre-compensated for dispersion and polarization-controlled with a half-wave plate (HWP). A dichroic beam splitter (DBS) transmits the fundamental light to a reflecting objective ($NA=0.5$), which focuses the light onto the device under test (DUT). The reflected 400~nm SHG signal is reflected by the DBS, filtered through a polarizer (Pol), and detected by a hybrid photodetector and a time-correlated single-photon counting (TCSPC) module.}
    \label{fig:setup}
\end{figure}

The experimental setup used to perform electric-field-induced second harmonic (EFISH) microscopy is shown schematically in Fig.~\ref{fig:setup}. A mode-locked Ti:sapphire oscillator (Spectra-Physics Element 2) generates fundamental laser pulses centered at $\lambda = 800$~nm with a pulse width of $\approx 10$~fs and a repetition rate of 80~MHz. To compensate for the group velocity dispersion introduced by the downstream optical components, a chirped-mirror dispersion compensation module is placed in the beam path. The beam then passes through a half-wave plate (HWP) to control the linear polarization state of the fundamental excitation. A dichroic beam splitter (DCBS) reflects wavelengths below 500~nm and transmits the 800~nm fundamental beam towards a reflecting objective lens ($NA = 0.5$). The reflecting objective focuses the fundamental light onto the reconfigurable \laosto\ nanojunction (the device under test, DUT) and collects the reflected signal. The second harmonic generation (SHG) signal generated at the junction at $\lambda = 400$~nm is collected in reflection, reflected by the DCBS, filtered through an analyzer (LP) to select the collected polarization component, and finally detected by a hybrid photodetector module coupled to a time-correlated single-photon counting (TCSPC) board (Becker \& Hickl). The sample is mounted in a Montana Instruments Cryostation for temperature control down to 6~K, with temperature stabilized to $\pm 0.5$~K.

\section{Power Characteristics}
To confirm the second-order optical nature of the nonlinear process at the nanojunction, the SHG intensity was measured as a function of the fundamental pump power under applied DC biases of $+1$~V and $-1$~V. The pump power was measured at the output of the first optical enclosure (referred to as the raw pump power). Due to reflection and absorption losses in the microscope optical path, the transmission efficiency to the device under test was calculated to be $13\%$. Thus, raw pump power values spanning 10--30~mW correspond to an actual incident power of 1.3--3.9~mW at the device.

\begin{figure}
    \centering
    \includegraphics[width=0.5\linewidth]{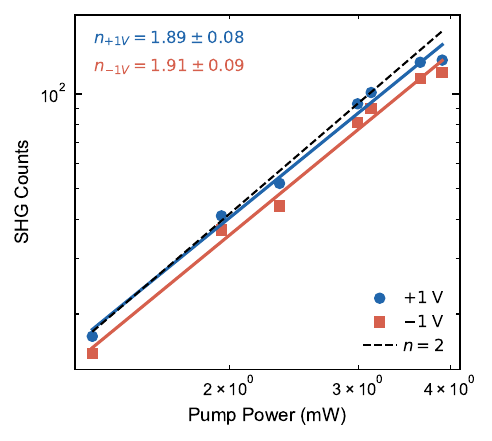}
    \caption{Power dependence of the second harmonic generation (SHG) signal at the reconfigurable \laosto\ nanojunction. SHG counts are plotted as a function of the incident pump power at the device for $+1$~V (blue circles) and $-1$~V (red squares) applied DC bias. The power at the device is calculated from the raw power at the first enclosure using a conversion factor of $13\%$. Solid lines show power-law fits $I(2\omega) \propto P^n$, yielding exponents of $n = 1.89 \pm 0.08$ and $n = 1.91 \pm 0.09$ respectively, consistent with the expected quadratic scaling of $n = 2$ (dashed line).}
    \label{fig:power_dep}
\end{figure}

Figure~\ref{fig:power_dep} shows the SHG intensity (photon counts) as a function of the incident pump power at the device for both $+1$~V and $-1$~V DC bias. The data is fitted to a power-law dependence of the form $I(2\omega) \propto P^n$. The fits yield power-law exponents of $n = 1.89 \pm 0.08$ at $+1$~V bias and $n = 1.91 \pm 0.09$ at $-1$~V bias. These exponents are close to the expected value of $n = 2$ for a second-order optical process (dashed black line in Fig.~\ref{fig:power_dep}), confirming that the detected signal is indeed second harmonic generation. The slight deviation from a perfect quadratic scaling may arise from minor beam alignment drifts.

\section{Absolute Susceptibility Calibration}
\label{si:calibration}

\subsection{Reference-crystal method}

The absolute magnitude of $\chi^{(3)}$ was determined by comparing the
device SHG rate to that of a BBO reference crystal under matched
excitation conditions (same wavelength, repetition rate, pulse
duration, and objective), following the standard comparative
(Maker-fringe-style) calibration approach~\cite{Herman1995-vx}. Both
signals were collected in reflection; no backing mirror was used for
either measurement.

\subsection{Backward-generation geometry}

Because both the device junction and the BBO reference were measured
in back-reflection without a backing mirror, the relevant generation
length is not the physical crystal thickness but the coherence length
for \emph{backward}-propagating second-harmonic generation. For a
pump propagating in $+z$, the nonlinear polarization at $2\omega$
is phased to $2k_\omega$; matching this source to a backward-going
free wave (wavevector $-k_{2\omega}$) requires
\begin{equation}
  \Delta k_\mathrm{back} = 2k_\omega + k_{2\omega}
  = \frac{2\omega}{c}\left(n_\omega + n_{2\omega}\right),
\end{equation}
in contrast to the forward-phase-matching mismatch
$\Delta k_\mathrm{fwd} = (2\omega/c)(n_{2\omega}-n_\omega)$. The
corresponding coherence length,
\begin{equation}
  \ell_c^\mathrm{back} = \frac{\lambda_\omega}{4\,(n_\omega+n_{2\omega})},
\end{equation}
sets the effective interaction length for reflected SHG, independent
of the physical crystal thickness once the crystal is thicker than
$\ell_c^\mathrm{back}$ (satisfied for our 10-$\mu$m BBO reference).
Using $n_\omega \approx 1.66$, $n_{2\omega}\approx1.68$ for BBO at
800/400~nm, $\ell_c^\mathrm{back}(\mathrm{BBO}) \approx 60$~nm. This
result follows the boundary-value treatment of
Bloembergen and Pershan~\cite{Bloembergen1962-wu}.

\subsection{Power-dependence calibration}
\label{si:calibration:bbopowerdep}

\begin{figure}
    \centering
    \includegraphics[width=0.5\linewidth]{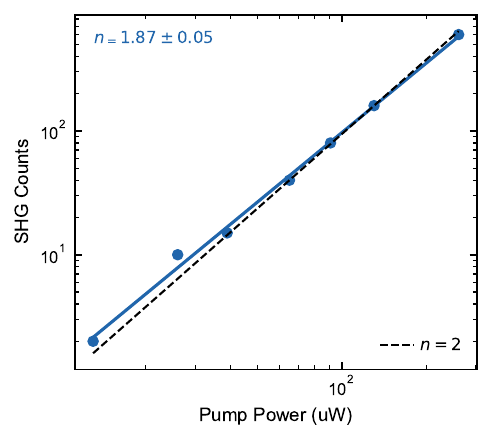}
    \caption{Power dependence of the BBO reference-crystal SHG signal used for absolute $\chi^{(3)}$ calibration. SHG counts from the 10-$\mu$m-thick BBO crystal, measured in back-reflection at room temperature under excitation conditions matched to the device measurement (same wavelength, repetition rate, and objective), are plotted as a function of average incident power. The solid line is a fit to $S(P) = kP^2$; confirming the quadratic scaling expected for second harmonic generation. The fitted rate at the reference power matching the device measurement (2~mW) is $I_\mathrm{BBO} \approx 3\times10^2$~counts/s, used in the $\chi^{(3)}$ extraction of Sec.~\ref{si:extracting}.}
    \label{fig:si_bbo_power}
\end{figure}

The BBO SHG rate was measured as a function of average incident
power (Fig.~\ref{fig:si_bbo_power}) and fit to $S(P) = k P^2$,
confirmed by the quadratic scaling of $S$ across the measured
range. At the reference power matching the device measurement
(2~mW), the fitted BBO rate was
$I_\mathrm{BBO} \approx 3\times 10^2$~counts/s.

\subsection{Extracting $\chi^{(3)}$ and $\chi^{(2)}_\mathrm{0}$}
\label{si:extracting}

Writing the device EFISH rate in terms of the fitted bias-sweep
coefficients ($I = a + bV + cV^2$; see main text Eq.~2), the pure
field-induced (quadratic) term at $V=1$~V is $c \approx 84$~counts/s.
Comparing to the BBO reference via the standard ratio expression,
\begin{equation}
  \chi^{(3)}_\mathrm{device}\,E_\mathrm{DC} =
  d_\mathrm{eff}^\mathrm{BBO}\,
  \frac{\ell_c^\mathrm{back}(\mathrm{BBO})}{\ell_\mathrm{device}}\,
  \sqrt{\frac{c}{I_\mathrm{BBO}}}\;
  \sqrt{\frac{n_\omega^2 n_{2\omega}\,(\mathrm{STO})}
             {n_\omega^2 n_{2\omega}\,(\mathrm{BBO})}},
\end{equation}
with $d_\mathrm{eff}^\mathrm{BBO}\approx2$~pm/V,
$\ell_\mathrm{device}\approx20$~nm (the junction gap),
and $E_\mathrm{DC}=5\times10^7$~V/m at $V_\mathrm{DC}=1$~V, gives
\begin{equation}
  |\chi^{(3)}_\mathrm{device}| \approx 1\times10^{-19}\ \mathrm{m^2/V^2}.
\end{equation}
This value carries a systematic uncertainty beyond ordinary
statistical scatter, discussed in Sec.~\ref{si:systematic} below.
The same procedure applied to the fitted zero-bias term $a$ gives
$|\chi^{(2)}_\mathrm{0}| \sim 0.1$--$1$~pm/V, consistent with the
nonzero $V_\mathrm{min}$ reported in the main text.

\subsection{Systematic uncertainty}
\label{si:systematic}

The BBO reference was measured at room temperature, while the device
measurement was made with the sample mounted in a closed-cycle
cryostation; both measurements used the same collection optics (the
same two dichroic beamsplitters, 500~nm shortpass filter, and 400~nm
bandpass filter). We did not independently verify that the overall
system throughput is otherwise identical between the two
measurements -- for example, the room-temperature versus
cryostat-mounted configurations could differ subtly in beam alignment
or focus -- and this comparison therefore carries a systematic
uncertainty beyond the statistical spread among the BBO power-sweep
data used to fit $S(P)=kP^2$ (Sec.~\ref{si:calibration:bbopowerdep}). We
therefore regard the quoted value, $|\chi^{(3)}_\mathrm{device}|
\approx 1\times10^{-19}$~m$^2$/V$^2$ (and the corresponding
$\chi_{1111}$, Sec.~\ref{si:chi1111}), as an order-of-magnitude
estimate rather than a precision measurement, with an estimated
systematic uncertainty of roughly a factor of two arising from this
potential mismatch.

\subsection{Relative phase between $\chi^{(2)}_\mathrm{0}$ and $\chi^{(3)}$}
\label{si:phase}

For real-valued $\chi^{(2)}_\mathrm{0}$ and $\chi^{(3)}$, the fitted
coefficients $a,b,c$ must satisfy $b^2 = 4ac$ exactly, since
$b = 2\chi^{(2)}_\mathrm{0}\chi^{(3)}$. Our fitted values
($a=22$, $b=8$, $c=84$, in counts/s) give $b^2 = 64$ versus
$4ac \approx 7390$, a substantial discrepancy. This is resolved by
allowing a relative phase $\Delta\phi$ between the two contributions,
$b = 2|\chi^{(2)}_\mathrm{0}||\chi^{(3)}|\cos\Delta\phi$, giving
$\cos\Delta\phi \approx 0.09$, i.e.\ $\Delta\phi \approx 85^\circ$.
This is physically reasonable given that $\chi^{(2)}_\mathrm{0}$
and $\chi^{(3)}$ are argued in the main text to arise from distinct
microscopic mechanisms (structural/defect-dipole symmetry breaking
versus soft-mode field response) with no requirement that they
respond in phase. One consequence is that $V_\mathrm{min}=-b/2c$
(main text Eq.~3) measures the in-phase projection of
$\chi^{(2)}_\mathrm{0}/\chi^{(3)}$ and should be treated as a lower
bound on the true magnitude ratio. Because this cross-term is small relative to the dominant quadratic
(EFISH) response, its temperature dependence, and thus that of
$\Delta\phi$ or $\chi^{(2)}_\mathrm{0}$ individually, is not resolved
by the present data (main text, Fig.~4).

\subsection{Isolating $\chi_{1111}$ from the bias-sweep geometry}
\label{si:chi1111}

The bias-sweep measurement of Sec.~S1 was performed with the fundamental
polarization set parallel to the junction axis, i.e.\ along the
crystallographic $[100]$ direction, which coincides with the direction
of $E_\mathrm{DC}$. Away from the biased junction \sto\ retains cubic
($m\bar{3}m$) symmetry, for which the third-order susceptibility
tensor $\chi^{(3)}_{ijkl}$, subject to the intrinsic permutation
symmetry $\chi^{(3)}_{ijkl}=\chi^{(3)}_{ikjl}$ appropriate to the two
degenerate optical legs, reduces to three independent components:
$\chi_{1111}=\chi_{2222}=\chi_{3333}$, $\chi_{1122}=\chi_{1212}$, and
$\chi_{1221}$.

With $E_\mathrm{DC}$ and the fundamental field both along
$\hat{x}\equiv[100]$, only $\chi_{ijkl}$ with an even number of $x$
and $y$ indices survive by the mirror symmetries of the point group;
setting $E_{\omega,y}=0$ in $P_i \propto \sum_{jkl}\chi_{ijkl}E_{\omega,j}E_{\omega,k}E_{\mathrm{DC},l}$
then gives $P_x(2\omega) = \chi_{1111}E_\omega^2$ and $P_y(2\omega)=0$:
the induced polarization is generated purely along the field/junction
axis, with no admixture of $\chi_{1122}$ or $\chi_{1221}$.

The detection analyzer, however, is oriented along the laboratory
$Y$-axis, which is rotated by $45^\circ$ from $[100]$ in our mounting
geometry (main text Fig.~3a). The detected amplitude is therefore the
projection $P_x/\sqrt{2}$, and the detected intensity is reduced by a
factor of two relative to the response along $[100]$ itself. The
value $\chi^{(3)}_\mathrm{device}$ obtained in Sec.~\ref{si:extracting}
is accordingly related to $\chi_{1111}$ by
\begin{equation}
  \chi_{1111} = \chi_{3333} = \sqrt{2}\,\chi^{(3)}_\mathrm{device}
  \approx 1.4\times10^{-19}\ \mathrm{m^2/V^2}.
\end{equation}
Because the fundamental field, $E_\mathrm{DC}$, and the junction axis
coincide with a single crystal axis in this measurement, this value is
not an orientation-averaged combination of tensor components: the
projection onto the fixed analyzer introduces only the geometric
factor above, and $\chi_{1122}$ and $\chi_{1221}$ do not enter. This
value likewise carries the order-of-magnitude systematic uncertainty
described in Sec.~\ref{si:systematic}.

\subsection{Constraint on $\chi_{1122}$ and $\chi_{1221}$ from the polarization dependence}
\label{si:polbound}

The polarization-resolved data of main text Fig.~3(b) provide an
independent check on the remaining two tensor components. For fixed
$E_\mathrm{DC}\parallel[100]$ and fundamental polarization rotated
through angle $\theta$ (measured from the laboratory $Y$-axis), the
analyzer-projected second-harmonic intensity contains, in addition to
a constant term, twofold and fourfold harmonics in $\theta$ with
amplitudes
\begin{align}
  a_2 &= -\tfrac{1}{2}\chi_{1122}(\chi_{1111}+\chi_{1221}), \\
  b_2 &= \tfrac{1}{4}(\chi_{1111}^2-\chi_{1221}^2), \\
  a_4 &= -\tfrac{1}{16}(\chi_{1111}-\chi_{1221})^2+\tfrac{1}{4}\chi_{1122}^2, \\
  b_4 &= -\tfrac{1}{4}\chi_{1122}(\chi_{1111}-\chi_{1221}),
\end{align}
up to an overall normalization. A fourfold ($\cos4\theta$,
$\sin4\theta$) component is present for any nonzero $\chi_{1122}$ or
$\chi_{1111}-\chi_{1221}$, and vanishes only in the limit
$\chi_{1111}\to\chi_{1221}$ and $\chi_{1122}\to0$ simultaneously. Our
polarization scans (Fig.~3b) show a clean two-lobed pattern, well
described by the twofold term alone, with no resolvable four-lobed
structure above the noise floor. Taking the background level of
$\sim$5~counts as a rough estimate of the smallest four-lobe amplitude
that would have been resolved, against the observed two-lobe amplitude
of $\sim$40~counts, the expressions above imply
$\chi_{1111}\approx\chi_{1221}$ and $|\chi_{1122}|\lesssim0.5\,\chi_{1111}$.
We regard this bound as order-of-magnitude, owing to the coarseness of the noise estimate, but it indicates that $\chi_{1111}$ and $\chi_{1221}$ are comparable in magnitude and jointly dominate the field-induced third-order response in this geometry, rather than $\chi_{1111}$ alone.

\section{Sample Growth, Device Fabrication, and Bias Application}
\label{si:fabrication}

\subsection{Sample growth}
LAO films of 3.4 unit cell thickness were deposited at 780~$^\circ$C on
STO in an oxygen partial pressure of $\sim 10^{-6}$~mbar, monitored in
situ by reflection high-energy electron diffraction (RHEED). Following
deposition, the sample was annealed at 600~$^\circ$C in 300~mTorr of
oxygen for one hour and then cooled to room temperature under the same
oxygen pressure.

\subsection{cAFM lithography}
Nanostructures were defined using a commercial atomic force microscope
(Asylum Research MFP-3D). Conducting nanowires were written parallel to the $[100]$
crystallographic axis of the SrTiO$_3$ substrate by scanning the tip at
$+15$~V sample bias in a humid environment (relative humidity
$\sim$40\%). The insulating gap was created by a single-pass cut at
$-10$~V sample bias. All cAFM operations were performed at room
temperature and ambient conditions.

\subsection{DC bias application}
DC bias was applied across the junction using a NI4461 PXIe DAQ card.
Gold electrodes ($\sim$100~nm thick) were deposited by photolithography
and e-beam evaporation and contacted with wire-bonded leads.

\bibliography{references}

\end{document}